# OS Scheduling Algorithms for Improving the Performance of Multithreaded Workloads

Murthy Durbhakula

Indian Institute of Technology Hyderabad, India

cs15resch11013@iith.ac.in, murthy.durbhakula@gmail.com

*Abstract*—Major chip manufacturers have all introduced multicore microprocessors. Multi-socket systems built from these processors are used for running various server applications. However to the best of our knowledge current commercial operating systems are not optimized for multi-threaded workloads running on such servers. Cache-to-cache transfers and remote memory accesses impact the performance of such workloads. This paper presents a unified approach to optimizing OS scheduling algorithms for both cache-to-cache transfers and remote DRAM accesses that also takes cache affinity into account. By observing the patterns of local and remote cache-to-cache transfers as well as local and remote DRAM accesses for every thread in each scheduling quantum and applying different algorithms, we come up with a new schedule of threads for the next quantum taking cache affinity into account. This new schedule cuts down both remote cache-to-cache transfers and remote DRAM accesses for the next scheduling quantum and improves overall performance. We present two algorithms of varying complexity for optimizing cache-to-cache transfers. One of these is a new algorithm which is relatively simpler and performs better when combined with algorithms that optimize remote DRAM accesses. For optimizing remote DRAM accesses we present two algorithms. Though both algorithms differ in algorithmic complexity we find that for our workloads they perform equally well. We used three different synthetic workloads to evaluate these algorithms. We also performed sensitivity analysis with respect to varying remote cache-to-cache transfer latency and remote DRAM latency. We show that these algorithms can cut down overall latency by up to 16.79% depending on the algorithm used.

# 1 INTRODUCTION

Many commercial server applications today run on cache coherent NUMA (ccNUMA) based multi-socket multi-core servers. Performance of multithreaded applications running on such servers are impacted by both cache-to-cache transfers and remote DRAM accesses. One way to alleviate this problem is to rewrite the application. Alternatively the operating system scheduler can be optimized to reduce the impact of both cache-to-cache transfers and remote DRAM accesses. In this paper we present a unified approach to optimizing OS scheduler for both cache-to-cache transfers and remote memory accesses that also takes cache affinity into account. In order to schedule a group of threads on a particular socket we need to take three factors into account: 1) Cache-to-cache transfers ii) Remote DRAM accesses and iii) Cache affinity. These three factors do not always align with each other and hence forces us to make certain tradeoffs. In this paper we present various scheduling algorithms that differ from each other in how they make these tradeoffs. They also differ from each other in complexity and benefit.

Section 2 discusses the scheduling algorithms. Section 3 describes the methodology I used in evaluating the scheduling algorithms. Section 4 presents results. Section 5 describes related work and Section 6 presents conclusions.

# 2 SCHEDULING ALGORITHMS

The general approach in all these algorithms is to first find optimal pairing of threads for reduced cache-to-cache transfers and then optimize that pair for reduced remote DRAM accesses.

**2.1 Algorithms for Optimizing Cache-To-Cache Transfers**

**2.1.1 Algorithm1**

In this algorithm for every thread i we find top three threads j,k,l with whom thread i has maximum cache-to-cache transfers. We then find a thread p which has highest cache-to-cache transfers among maximum pairs (i,j). Then we pick top three threads j,k,l with whom thread p has maximum cache-to-cache transfers. This forms our first group of four threads. We then remove p,j,k,l from the 16 threads. We repeat above algorithm with remaining 12 threads and get our second group of four threads. We then remove second group of four threads from 12 threads. We continue above process till we form all four groups of four threads.

We can see that in the above algorithm we have run the routine of finding maximum in N threads a constant number of times where the constant number is dependent on number of sockets. Hence the complexity of above algorithm is : O(NKL) where N is total number of threads, K is number of threads per socket, and L is total number of sockets. Below is high-level pseudo-code of this algorithm.

**Input:** Threads T0,…TN with cache-to-cache transfer among themselves. Every pair of threads can have cache-to-cache transfers. Hence there are Nc2 cache-to-cache transfers. Each node has 4 cores and can run one thread per core. Current schedule S_current which has a grouping of N threads divided into L groups. Each group can run on one node.

**Output**: A new schedule S_next with new grouping of threads for optimized cache-to-cache transfer latency

begin

1. For every thread i, find a thread j with which it has maximum cache-to-cache transfers. Let's call the maximum cache-to-cache transfers as Max-i. This can be done in 0(N) time.
2. Find a thread p which has highest maximum cache-to-cache transfer Max-p among all the N maximum cache-to-cache transfers; one per thread. This can also be done in 0(N) time.
3. Pick thread p and its pair j with which it has maximum cache-to-cache transfers. Now find other threads k, l with which thread p has second and third highest cache-to-cache transfers. These can also be done in 0(N) time. Threads (p,j,k,l) forms the first optimal group of threads in S_next. Remove this group from N threads and repeat steps 2 and 3 until all L groups are formed.

end

**Overall complexity of the algorithm:** Step 1 and 2 have 0(N) complexity. Step 3 has O(N(K-1)) where K is number of threads per core. Steps 2 and 3 are run L times where L is number of nodes. Hence overall complexity is 0(NKL).

<div align="center">**Algorithm 1 for Cache-to-Cache Transfer Optimization**</div>

### 2.1.2 Algorithm2

In this algorithm we sort pair of cache-to-cache transfers. There are Nc2 such pairs. Then we find pair (i,j) which has highest cache-to-cache transfers. We then traverse the sorted list to find next highest pair (k,l). If neither k nor l matches (i,j) then (i,j,k,l) forms first group of four threads. If either k or l matches either of (i,j) then we do the following. Let us assume, without any loss of generality, that k is i and l does not match i or j. Then we find a thread p which has highest cache-to-cache transfers with thread l and does match i or j. Now (i,j,l,p) forms first group of four threads. Once we get first group of four threads we remove them from 16 threads. Then we run the above procedure for the remaining 12 threads to find second group of four threads and so on until we find all four groups of four threads.

The complexity of above algorithm is Nc2log(Nc2) because sorting of Nc2 pairs dominates the complexity of above algorithm. Below is high-level pseudo-code of this algorithm.

**Input:** Threads T0,…TN with cache-to-cache transfer among themselves. Every pair of threads can have cache-to-cache transfers. Hence there are Nc2 cache-to-cache transfers. Each node has 4 cores and can run one thread per core. Current schedule S_current which has a grouping of N threads divided into L groups. Each group can run on one node.

**Output:** A new schedule S_next with new grouping of threads for optimized cache-to-cache transfer latency

begin

1) Sort all Nc2 pair of cache-to-cache transfers in descending order. This can be done in O(Nc2log(Nc2)) time.
2) Start from highest pair (i,j). Traverse the sorted pairs in descending order. As you traverse find next highest pair (k,l). If (k,l) does not overlap with (i,j) then (i,j,k,l) forms first optimal group of threads in S_next. Else if ( k,l) overlaps then we find non-overlapping (l,p) as explained above. (i,j,l,p) forms first optimal grouping of four threads. Remove this group of threads from N threads and repeat Step 2 to find the next group of threads. This can be done in O(Nc2) time.

end

**Overall complexity of the algorithm is:** O(Nc2log(Nc2)).

## 2.2 Algorithms for Optimizing DRAM accesses

### 2.2.1 Algorithm1

In this algorithm, we first sort all local and remote memory access counts for each group of threads along with node information in monotonically decreasing order. We start from top and assign the thread-group with highest access count to that node. Say thread-group G0 has highest access count of 10000 to node 2 then we assign G0 to node 2. Then we go to next element and assign the next thread-group G1 to its corresponding node. Once we assign a thread-group to one node we cannot assign it to another node. So in the above example once G0 is assigned to node 2, it cannot be assigned to any other node. The complexity of the algorithm is of order $O((L*L)\log(L*L))$ where L is total number of thread groups as well as nodes. This is because the sorting algorithm dominates the complexity and we use merge-sort. There are specialized algorithms such as counting sort which could further reduce complexity, however their application is data dependent. Below is high-level pseudo-code of this algorithm.

**Input:** L groups of threads with DRAM accesses to L nodes. Each node has 4 cores and can run one thread per core. Current schedule S_current which has a mapping of L groups to L nodes. Each group has 4 threads.

**Output:** New schedule S_next with new mapping of groups to nodes.

begin

1. Each of L groups have L DRAM access counts; one per node. Sort L*L DRAM accesses in descending order. Complexity is $O((L*L)\log(L*L))$.

2. Scan DRAM access counts. Start from highest DRAM access count and say it is coming from group L1 to node N2. Assign L1 to N2 in schedule S_next. Now L1 cannot be assigned to any other node.

3. Repeat step 2 until all groups are assigned a node in S_next. Complexity is $O(L*L)$.

end

**Overall complexity of the algorithm is:** $O((L*L)\log(L*L))$.

### Algorithm 1 for Remote DRAM Access Optimization

### 2.2.2 Algorithm 2

In this algorithm we first start off with node 0 and we sort all thread-group's accesses to that node in monotonically decreasing order. Then we pick first thread-group with highest accesses to node 0. Then we remove that thread-group from the list. We then sort thread accesses to node 1 out of 3 remaining thread-groups. We pick next top thread-group to node 1. We remove that thread-group from the list. We sort accesses to node 2 from the remaining two thread-groups. We pick next top thread-group to node 2. Remaining thread-group will go to node 3. The complexity of this algorithm is order $O(L\log(L))$ assuming sorting is done in parallel on all nodes, where L is total number of thread-groups. Even here it is sorting that dominates the complexity. Below is high-level pseudo-code of this algorithm.

**Input:** L groups of threads with DRAM accesses to L nodes. Each node has 4 cores and can run one thread per core. Current schedule S_current which has a mapping of L groups to L nodes. Each group has 4 threads.

**Output:** New schedule S_next with new mapping of groups to nodes.

begin

1. Sort all DRAM accesses from L groups of threads in descending order. If all nodes do it in parallel, Complexity is O(Llog(L)). Steps 2 and 3 has to be done serially by all nodes involved.

2. Start with node N0 and pick top group of threads and assign them to N0 in schedule S_next. Now this group of threads cannot be assigned to any other node.

3. Repeat step 2 for all nodes until all groups of threads are assigned a node in S_next. Complexity is O(L).

end

**Overall complexity of the algorithm is:** O(Llog(L)).

**Algorithm 2 for Remote DRAM Access Optimization**

## 3 METHODOLOGY

We implemented each of these algorithms as a stand-alone C++ program and evaluated them by running synthesized data access patterns. These synthesized data access patterns vary both cache-to-cache transfers as well as local and remote dram access counts in a known pattern for each scheduling quantum. The data is then fed to each of the algorithms to see if they can track the pattern. The overall benefit for each algorithm depends on how well they can track the pattern. We chose synthetic workloads in order to clearly bring out benefits of each algorithm and its sensitivity to remote cache-to-cache transfer latency and remote DRAM latency. As part of future work we plan to study the impact of these algorithms on real workloads. In a real system we need performance counters to count number of cache-to-cache transfers between every pair of threads as well as performance counters to count local and remote DRAM accesses for every thread. This is the only hardware support we need for the scheduler optimizations.

Table 1 lists base configuration used for evaluation of different scheduling algorithms. On top of the base configuration we also perform sensitivity analysis of different algorithms with respect to varying remote cache-to-cache transfer latency from 100 to 175 to 250 cycles and varying remote DRAM latency from 250 to 375 to 500 cycles. Table 2 lists three synthetic cache-to-cache access patterns and Table 3 lists three synthetic DRAM access patterns used in this paper. A high level system diagram can be seen in Figure 1. The circled portion represents one socket with 4 cores per socket.

| Parameter | Value |
|---|---|
| CPU Frequency | 2 Ghz |
| Number of cores per socket | 4 |
| Number of sockets/nodes | 4 |
| Shared Cache | 1 MB |
| Local DRAM Latency | 125 cycles |
| Remote DRAM Latency | 250 cycles |
| Cache-to-Cache Latency within socket | 50 cycles |
| Cache-to-Cache Latency across sockets | 100 cycles |
| Number of scheduling Quanta | 16 |

**Table 1: Configuration Parameters**

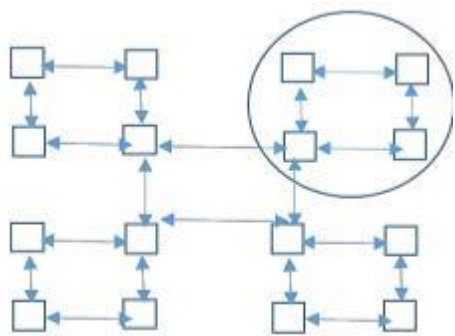

**Figure 1: High-level ccNUMA Multi-socket Multi-core System diagram with 4 sockets and 4 cores-per-socket**

| Access Pattern | Description |
| --- | --- |
| **Synth1_C_To_C - Single phase access pattern** | Same cache-to-cache access pattern for all quanta except first quanta. Single optimal grouping of threads. |
| **Synth2_C_To_C - Two phase access pattern** | Two cache-to-cache access patterns equally distributed over all quanta. Optimal grouping of threads from 1st till 8th quanta. Second optimal grouping of threads from 9th till 16th quanta. |
| **Synth3_C_To_C - Four phase access pattern** | Four cache-to-cache access patterns equally distributed over all quanta. Optimal grouping of threads from 1st till 4th quanta. Second optimal grouping of threads from 5th till 8th quanta. Third optimal grouping of threads from 9th till 12th quanta. Fourth optimal grouping of threads from 13th till 16th quanta. |

**Table 2: Synthetic workloads for cache-to-cache transfers**

| Access Pattern | Description |
| --- | --- |
| **Synth1_DRAM - Single phase access pattern** | Same DRAM access pattern for all quanta except first quanta. Single optimal grouping of threads. |
| **Synth2_DRAM - Two phase access pattern** | Two DRAM access patterns equally distributed over all quanta. Optimal grouping of threads from 1st till 8th quanta. Second optimal grouping of threads from 9th till 16th quanta. |
| **Synth3_DRAM - Four phase access pattern** | Four cache-to-cache access patterns equally distributed over all quanta. Optimal grouping of threads from 1st till 4th quanta. Second optimal grouping of threads from 5th till 8th quanta. Third optimal grouping of threads from 9th till 12th quanta. Fourth optimal grouping of threads from 13th till 16th quanta. |

**Table 3: Synthetic workloads for remote DRAM accesses**

## 4 RESULTS

In this section we present results of various experiments done on algorithms described in section 2 using synthetic workloads described in section 3. This work can be extended to evaluate these algorithms on real workloads by running similar experiments.

### 4.1 Only Cache-to-Cache Transfers Optimization

First we run the three synthetic workloads Synth1_C_To_C-Synth1_DRAM, Synth2_C_To_C-Synth2_DRAM, and Synth3_C_To_C-Synth3_DRAM with only cache-to-cache transfer optimization algorithms. Those results are plotted in the graph below. The graph shows percentage improvement in overall latency.

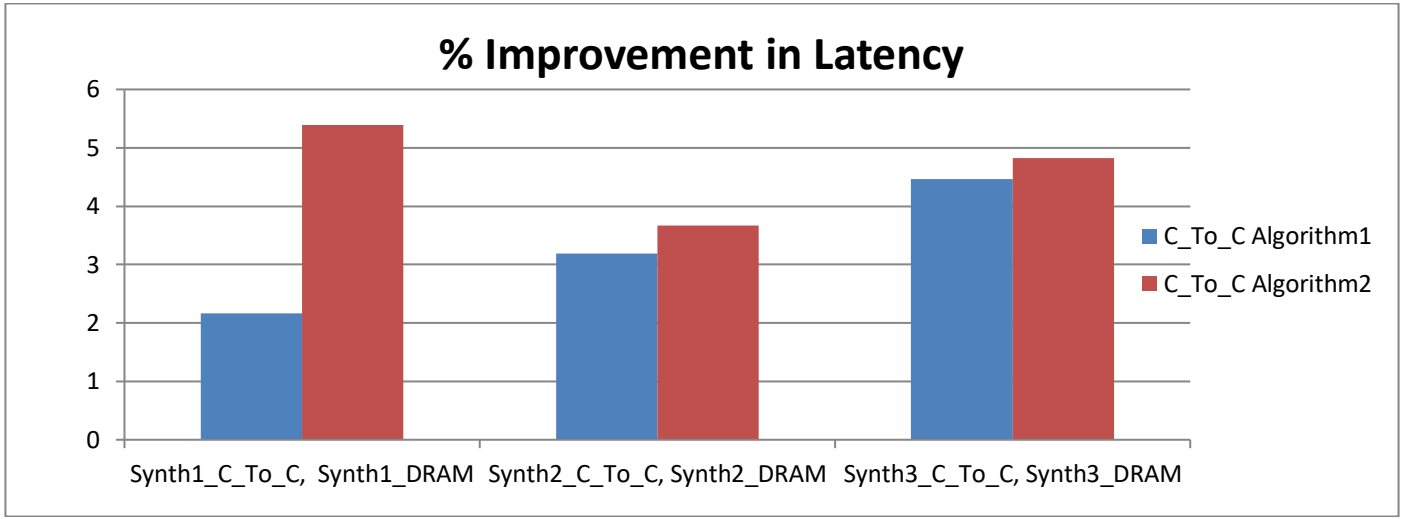

Figure 2: Percentage improvement in overall latency with only cache-to-cache transfer optimization

associated with each algorithm. As we can see Algorithm 2 consistently outperforms Algorithm 1. However the extra improvement we get with Algorithm 2 is not always worth the high algorithmic complexity associated with Algorithm 2. We also observe that sometimes performance improves as we move from Synth1_C_To_C- Synth1DRAM to Synth2_C_To_C-Synth2 DRAM or from Synth2_C_To_C-Synth2DRAM to Synth3_C_To_C -Synth3DRAMand sometimes performance decreases. On further analysis we found that this is dependent on whether we get more or less benefit when there is a phase change in the workload. Sometimes due to phase change performance improves and sometimes performance decreases. It is workload and algorithm dependent. However, irrespective of workload, Algorithm 2 outperforms Algorithm 1 when only cache-to-cache transfer optimization is enabled.

## 4.2 Both Optimizations together without taking Cache Affinity into Consideration

In this section we present results of running the synthetic workloads when both cache-to-cache transfer optimization and remote DRAM optimizations are on while not taking cache-affinity into account.

**Note:** For rest of this paper, for remote DRAM optimization, we are only showing Algorithm 2 because we did not find any major difference in results between Algorithm 2 and Algorithm 1 and Algorithm 2 is of lower complexity than Algorithm1.

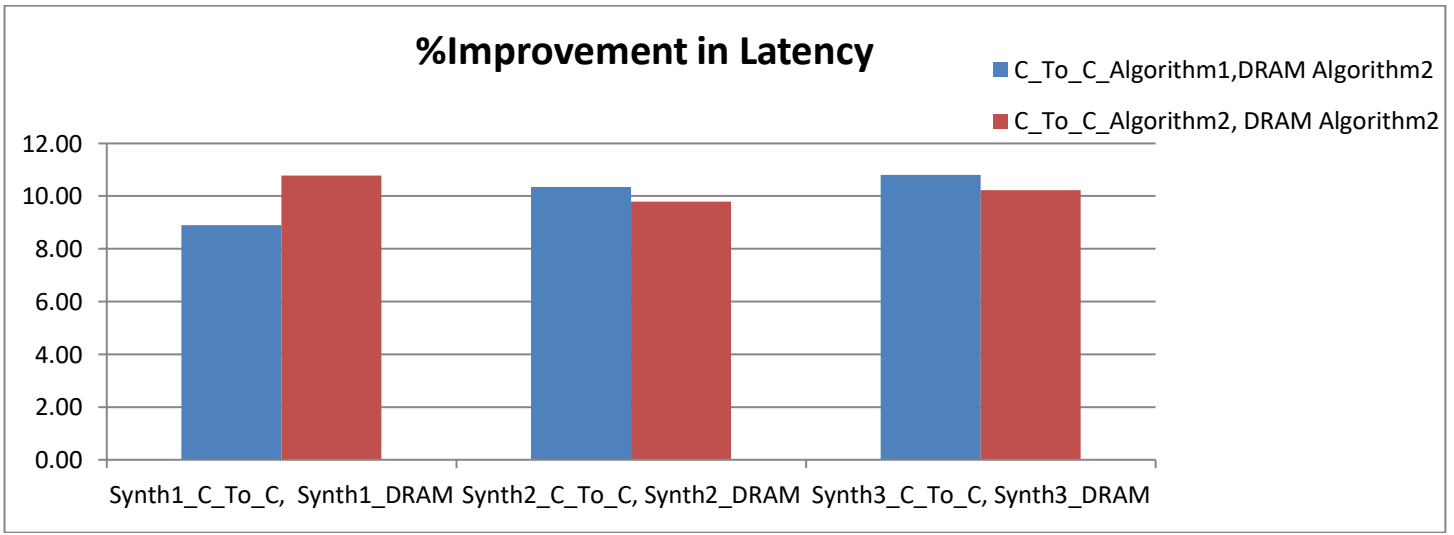

Figure 3: Percentage improvement in overall latency with both optimizations without Cache Affinity

We can see from the above graph that with remote DRAM optimization on sometimes C_To_C_Algorithm1 outperforms C_To_C_Algorithm2. This is because remote DRAM latencies are typically higher than remote cache-to-cache latencies. However one key takeaway is that with remote DRAM optimization added percentage improvement in latency is much higher than percentage improvement in latency with only cache-to-cache transfer optimization on, as shown in Figure 2.

### 4.3 Both Optimizations together with Cache Affinity

In this section we present results of running the synthetic workloads when both cache-to-cache transfer optimization and remote DRAM optimizations are on while taking cache-affinity into account. Whenever a thread shifts from its current socket to another socket we associate a penalty due to cache affinity. In each socket four threads share a 1MB cache. Hence roughly each thread can use 256 KB of data. Of which we assume that in the next quanta when it gets scheduled it still can use 64 KB of worthwhile data. We experimented with various fractions of 256 KB data and concluded that 64 KB is a reasonable choice. Hence, assuming cache line size of 64 bytes, cache transfer penalty is same as transferring (1024) cache lines to another socket. Assuming remote cache transfer latency of 250 cycles, it is 250*1024 = 256000 cycles.

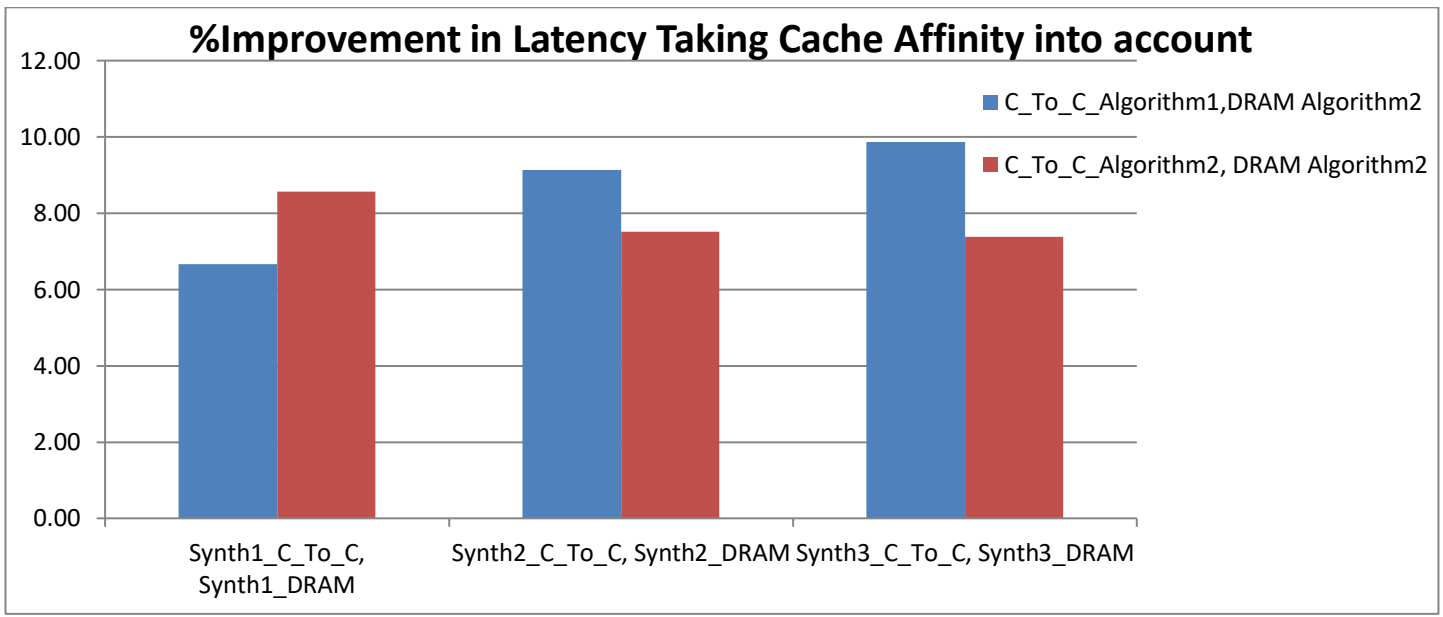

**Figure 4: Percentage improvement in overall latency with both optimizations with Cache Affinity**

We can see from the above graph that with cache affinity the benefit due to combined cache-to-cache transfer latency and remote DRAM latency optimization has reduced. However it is still much better than applying only cache-to-cache transfer latency optimization. We can also observe that C_To_C Algorithm1 performance is very close to that of C_To_C Algorithm2, sometimes outperforming C_To_C Algorithm2. Hence given higher algorithmic complexity of C_To_C Algorithm2, C_To_C Algorithm1 is a better choice unless the number of threads N is small enough that it does not make much of a difference.

### 4.4 Sensitivity Analysis

In this section we show results of applying sensitivity analysis to combined algorithms with cache affinity while varying both remote cache-to-cache transfer latencies and remote DRAM latencies. We increased both remote cache-to-cache transfer latencies and remote DRAM latencies as follows:

**1) Experiment 1: Remote Cache-to-Cache Latency: 175 cycles, Remote DRAM Latency: 375 cycles**

Here we increased remote cache-to-cache transfer latency to 175 cycles keeping local cache-to-cache transfer latency unchanged. Further we have also increased remote DRAM latency to 375 cycles keeping local DRAM latency unchanged.

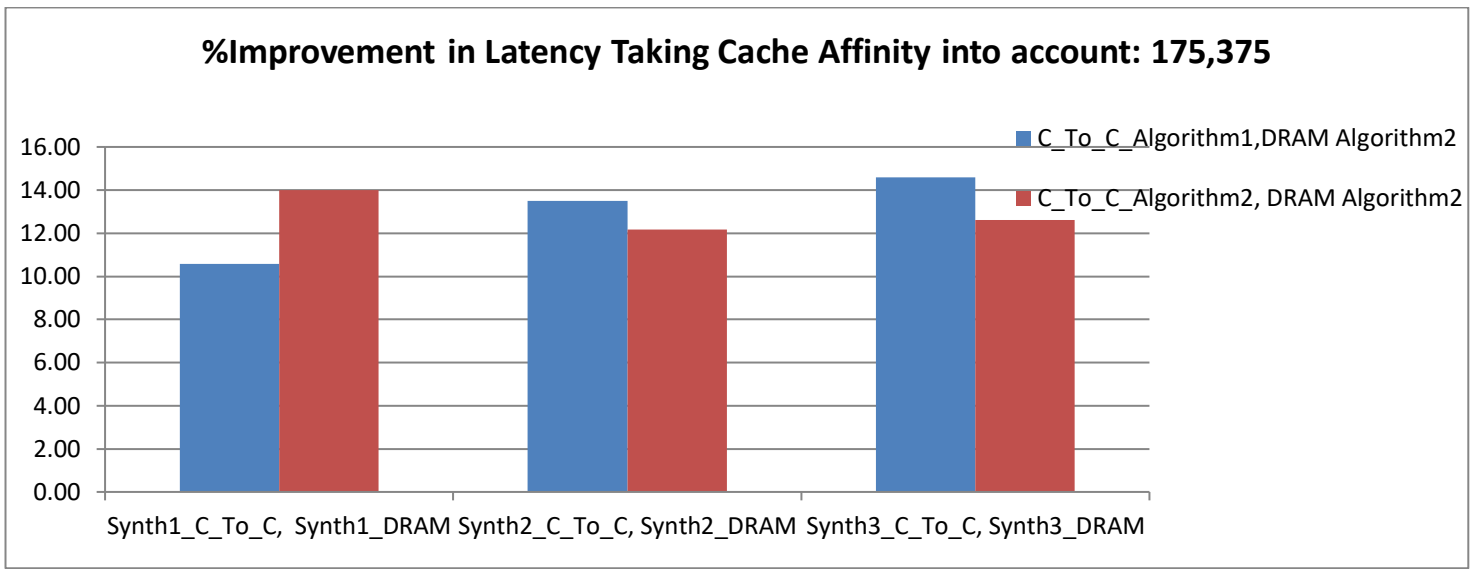

**Figure 5: Percentage improvement in overall latency with both optimizations and cache affinity - Sensitivity analysis: 175,375**

We can see from the above graph that performance improves with increased remote cache-to-cache transfer latency and increased remote DRAM latency. This is to be expected since the algorithms are trying to reduce the number of remote cache-to-cache transfers and the number of remote DRAM accesses.

**2) Experiment 2: Remote Cache-to-Cache Latency: 250 cycles, Remote DRAM Latency: 500 cycles**

Here we increased remote cache-to-cache transfer latency to 250 cycles keeping local cache-to-cache transfer latency unchanged. Further we have also increased remote DRAM latency to 500 cycles keeping local DRAM latency unchanged.

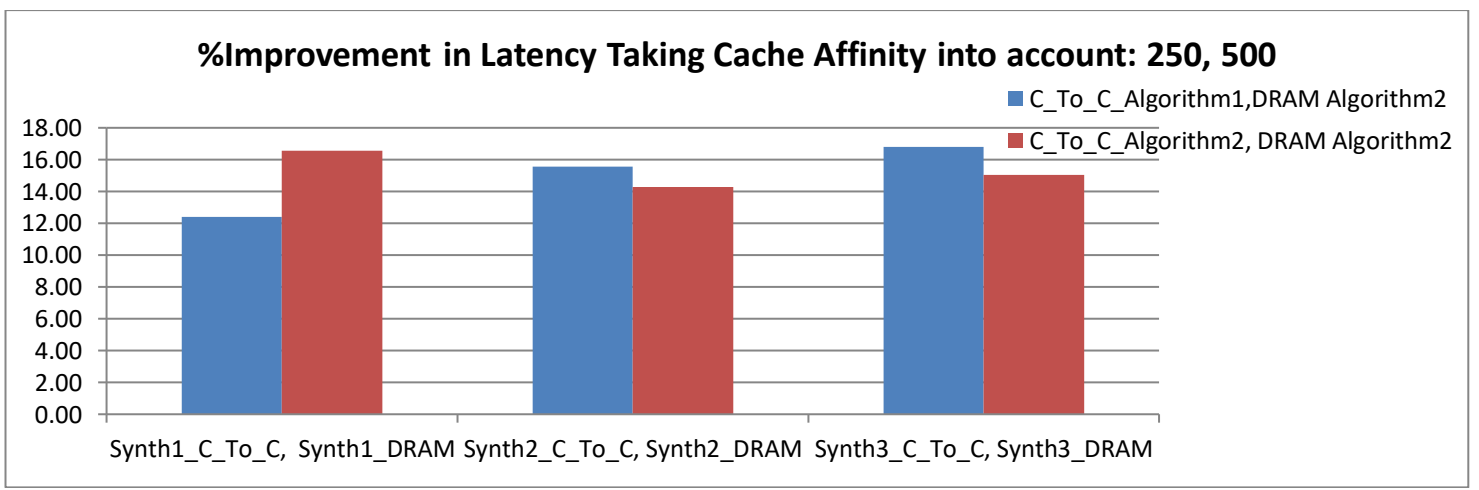

**Figure 6: Percentage improvement in overall latency with both optimizations and cache affinity - Sensitivity analysis: 250, 500**

As expected we can see from the above graph that the performance improves even further compared to experiment 1. Again this is because the algorithms are trying to reduce the number of remote cache-to-cache transfers and the number of remote DRAM accesses.

## 5  RELATED WORK

To the best of my knowledge this is the first study which presents a unified approach to optimizing both cache-to-cache transfers and remote DRAM accesses while taking cache affinity into account. There have been studies which tackled each of these problems separately which we discuss below.

Thread-clustering algorithms were examined by Thekkath and Eggers[1]. Their research dealt with finding the best way to group threads that share memory regions together onto the same processor so as to maximize cache sharing and reuse. However, the way they track sharing is address-based and can be prone to errors, particularly for migratory sharing patterns. If processes P0, P1, P2 are involved in migratory sharing, and if you select P0 and P1 to schedule on the same socket, and if the cache line actually migrates from P0 to P2 and then back to P1 then we would not have reduced inter-socket cache-to-cache transfers. In our work we track sharing using pair-wise counters. We also present a better global algorithm. In our work, we assume that the number of threads matches the number of processors: in other words, the system is always load-balanced.

Our work most closely resembles the work done by Tam et al [2]. Their paper also proposes using hardware performance counters for thread clustering in a shared-memory multiprocessor system. We use a different approach to collect inter-socket cache-to - cache transfers using our performance counters. We do not count remote cache stalls per address range (since it can be prone to the migratory sharing problem described above); instead, we count pair-wise remote cache transfers. This approach works for small and medium-range servers. For bigger machines (such as Cray Red Storm, Aprro International Atlas [3], etc) we can use a coarse - grained approach such as combining close-by sockets as one "logical socket". This is so that, even if w e may not schedule highly interacting threads on the same socket we will schedule them on close-by sockets to reduce the number of hops. We assume that the home node for all cache-to-cache transfers for the selected cluster of threads falls in the same socket. One way to overcome this problem is to maintain a per-home-node-based counter for counting cache-to-cache transfers between every pair of threads, and then take the home node into consideration for the scheduling algorithm. Sridharan *et al*. [4] examined a technique to detect user space lock sharing among multi-threaded applications by annotating user-level synchronization libraries. Using this information, threads sharing the same highly contended lock are migrated onto the same processor. Our work adopts the same spirit but at a more general level that is applicable to any kind of memory region sharing.

Many researchers have investigated OS support for minimizing cache conflict and capacity problems of shared L2/L3 cache processors [5,6,7]. Our work, however, focuses on reducing the impact of communication misses that are inherent to applications. Our work may be complementary to these past efforts.

Chandra el al.[8] have studied impact of various OS scheduling policies on performance of both uniprocessor and multiprocessor workloads. Our work is similar to theirs in that we also study OS scheduling algorithms for improving performance of parallel workloads. However our focus more on memory intensive workloads for ccNUMA multi-socket multi-core servers. We assume that at-a-time only one multi-threaded application is running with each core assigned one thread for the sake of load-balancing. The algorithms we propose are completely different. Their work also evaluated benefits of page migration. However page migration & replication are not always beneficial. For instance say thread T0 is scheduled on Node 1 with 1000 accesses to Node 1's DRAM and also accesses a page P2 on Node 2 DRAM, with intention-to-write, 500 times. At the same time thread T1 is scheduled on Node 3 with 2000 accesses to Node 3's DRAM and accesses same page P2 on Node 2 DRAM, with intention-to-write, 500 times. In such a situation it is not possible to decide to which node to migrate P2 to. In fact if there are such "hot" pages accessed by multiple threads it is actually better to schedule them on the same node where hot page is, rather than migrating the hot page.

Kaseridis et al. [9] proposed a dynamic memory subsystem resource management scheme that considers both cache capacity and memory bandwidth contention in large multi-chip CMP systems. Their memory bandwidth contention algorithms monitor when a particular schedule of threads exceed the maximum bandwidth supported by a node and then try to schedule bandwidth demanding threads with those threads that need little memory bandwidth. Whereas our approach proactively tries to find the best pairing of threads for any scheduling quanta while keeping the overall bandwidth utilization within the maximum bandwidth limits. And the algorithms presented in this paper are different from their algorithms.

Ipek et al [10] proposed using reinforcement learning based approach to tune DRAM scheduling policies to effectively utilize off-chip DRAM bandwidth. Their work differs from ours in two different ways. First, we are using OS scheduling algorithms to optimize DRAM bandwidth utilization. Second, we use parallel workloads to evaluate the scheduling policy where as they use multiprogramming workloads with no sharing.

Ahn et al [11] studied the impact of DRAM organization on the performance of data parallel memory systems. In contrast to their work we focus on novel OS scheduling algorithms that will improve DRAM performance of parallel applications on general purpose multiprocessors. Zhu et al [12] proposed using novel DRAM scheduling algorithms for SMT processors. In contrast to their work this paper proposes using new OS scheduling algorithms with minimal hardware support.

Tang et al.[14] studied the impact of co-locating threads of different multi-threaded applications in a data-center environment on overall performance. They proposed heuristics for co-locating different workloads. Their focus is mostly on data-center related workloads and the algorithms presented in this work are different from their heuristics.

More recently Kim and Huh [17] proposed OS scheduling algorithms for implementing fairness in multi-core systems. They proposed a few algorithms that improve fairness in heterogeneous multi-core systems while improving overall throughput. Their focus is on mainly improving fairness using OS scheduling policies while the focus of this paper is on reducing remote cache-to-cache transfers and remote DRAM accesses thereby improving overall performance. Further Kim and Huh use SPEC workloads for the evaluation of their algorithms whereas we focus on parallel workloads.

Sahoo and Dehury [18] proposed job scheduling algorithms for CPU intensive workloads in a health-care cloud environment. Their focus is mainly on health-care applications in a cloud-based environment. Whereas the focus of this paper is on improving performance of memory intensive parallel workloads.

There are many other studies [13, 15, and 16] which focused on tuning DRAM scheduling policies or memory access ordering for better overall performance. Our work is different from these in two ways. First we focus on OS scheduling algorithms to reduce the impact of remote cache-to-cache transfers and remote DRAM accesses. Second, these studies [13] focus on optimal DRAM utilization for co-located single threaded workloads where as our work focuses on improving performance of a multi-threaded parallel workloads.

# 6 CONCLUSION

Many commercial server applications today run on cache coherent NUMA (ccNUMA) based multi-socket multi-core servers. Performance of multithreaded applications running on such servers are impacted by both cache-to-cache transfers and remote DRAM accesses. Often times optimizing for both these issues requires us to make certain tradeoffs. In this paper we have presented a unified approach to optimizing OS scheduling algorithm for both cache-to-cache transfers and remote DRAM accesses that also takes cache affinity into account.

We presented two algorithms of varying complexity for optimizing cache-to-cache transfers. One of these is a new algorithm which is relatively simpler and performs better when combined with algorithms that optimize remote DRAM accesses. For optimizing remote DRAM accesses we presented two algorithms. Though both algorithms differ in algorithmic complexity we find that for our workloads they perform equally well. We used three different synthetic workloads for both cache-to-cache transfers and DRAM accesses to evaluate these algorithms. Further we also performed sensitivity analysis with respect to varying remote cache-to-cache transfer latency and remote DRAM latency. We showed that these algorithms can cut down overall latency by up to 16.79% depending on the algorithm used.

This work could be extended in several ways. One way is to use machine learning algorithms to learn any kind of phase behavior among prior scheduling quanta and incorporate that into scheduling decision for the next quanta. Another way is to monitor the benefit of these algorithms every scheduling quanta and turn off/on the optimization adaptively if benefit falls  In general for any long running application with stable patterns, hardware could provide feedback to the OS, which could in turn use that information to adapt its policies to benefit application performance.

ACKNOWLEDGEMENTS

I would like to thank Prof. Alan Cox of Rice University for initially discussing with me the concept of optimizing OS scheduling algorithms for improving the performance of various workloads. And I would like to thank  my wife and kids for supporting me morally during the course of this work.